\documentclass[manuscript]{aastex}
\shorttitle{Time delay and microlensing in Q0142-100}
\shortauthors{Oscoz et al.}
\begin{document}
\title{Long-term monitoring, time delay and microlensing in the gravitational lens 
system Q0142-100}
\author{A. Oscoz\altaffilmark{1,2}, M. Serra-Ricart\altaffilmark{1,2}, E. 
Mediavilla\altaffilmark{1,2}, and J. A. Mu\~noz\altaffilmark{3}}

\altaffiltext{1}{Instituto de Astrof\'\i sica de Canarias, C. V\'\i a L\'actea s/n,
E-38205, La Laguna, Spain}
\altaffiltext{2}{Departamento de Astrof\'\i sica, Universidad de La Laguna, La Laguna, Spain}
\altaffiltext{3}{Departamento de Astronom\'\i a y Astrof\'\i sica, Universidad de Valencia, 
46100-Burjassot, Valencia, Spain}

\begin{abstract}
We present twelve years of monitoring of the gravitational lens Q0142-100 from the Teide
Observatory. The data, taken from 1999 to 2010, comprise 105 observing nights with 
the IAC80 telescope. The application of the $\delta^2$-method to the dataset leads to a 
value of the time delay between both components of the system of 72$\pm22$
days (68 per cent confidence level), consistent within the uncertainties with 
the latest previous results. 
With this value in mind a possible microlensing event is detected in Q0142-100.
\end{abstract}


\section{Introduction}

Q0142-100 (UM 673), discovered by \citet{mac82}, was classified as a gravitational lens 
system for 
the first time by \citet{sur87,sur88}. This double quasar shows two identical images 
($z = 2.719$) separated by 2.2 arcsecs with the lensing galaxy ($z = 0.49$) placed 
among them (\citet{sur88}, \citet{sme92}, \citet{eig07}), see figure 1. The specially 
propitious 
configuration of Q0142-100, with large separation between both images, and their magnitudes 
($m_R(A) \sim 16.5$ and $m_R(B) \sim 18.4$) make this system adequate for a photometric 
monitoring from a medium-sized telescope. Moreover, a predicted time delay of only some 
months should have turned Q0142-100 a very attractive target for time delay determinations. 
However, the difficulties coming from a lensing galaxy placed very close to the B, 
fainter, component, the low variations in magnitude of the light curves of both 
components and the large annual gaps in the monitoring make strongly difficult the 
calculation of the time delay. From Teide Observatory, for example, reasonably close to 
the equator, Q0142-100 can be observed about 250 days per year but this becomes in a 
210-220 days window each year when the mechanical limits of the telescope are considered.

This gravitational lens has been monitored since its discovery without very effective 
results. \citet{dau93} observed it in the $B$ band from 1987 to 1993 but their 
observations were of modest quality, concluding that no significant variability was 
seen at the quasar. Some years later, \citet{sin01} presented 29 $R$ band observations 
taken from 1995 to 2000, detecting variations of 0.2 mag in their data and a global 
variation of $\sim0.5$ mag since the system was discovered. \citet{nak05} performed 
two-band, $V$ (23 points) and $i$ (18 points), photometry during three years, October 
1998 to December 2001. Unfortunately, they were unable to measure the time delay due 
to the small variations in magnitude of the light curves, but found possible evidence 
for microlensing: component A, the brighter component, became bluer as it got brighter. 
Microlensing should be achromatic but for the inner parts of compact accretion disks
the size of the emitting region varies with wavelength and hence the microlensing 
magnification depends on the wavelength (see, e.g., \citet{med11} and references therein). 
\citet{nak05} results point to a microlensing of the compact accretion disk: the central, 
bluer, part of the source is more amplified than the outer, redder parts.

Recently, \citet{kop10} 
obtained an estimate of the time delay between both components of Q0142-100 from 
observations made in the $V$, $R$ and $I$ bands during the 2003-2005 period. These 
authors used a telescope placed at the Maidanak Observatory, and they were able to 
observe this lens only 3 months each year. They claimed that the 
time delay between A and B components of Q0142-100 is of about 150 days, with component 
B following the brightness variations of component A. To obtain this value the observations 
at different bands are combined and then they interpolate up to 120 days to fill on the 
large gaps in which Q0142-100 cannot be observed. Finally, \citet{kop12} mix these data with
archival data from the Maidanak Observatory, additional Maidanak observations and 2008-2010 
CTIO data. Their new database covers the period 2001-2010, ten observational seasons, 
and leads them to a new and lower value for the time 
delay: $95_{-16 -29}^{+5 +14}$ days (68 and 95\% confidence intervals), with 
$89\pm11$ days as the most probable result.

In this paper we present the results of the photometric follow-up of Q0142-100 made with the 
IAC80 telescope during more than ten years. This is, to our knowledge, one of the longest 
monitoring ever made of a gravitational lens system and comparable, of course in a lower
scale, to what LSST will do. This monitoring includes 105 
valid sessions of observation in the $R$ 
band, and has been done within the gravitational lens programme carried 
out at the Instituto de Astrof\'\i sica de Canarias. The observations and reduction 
processes are presented in Section 2, and the time delay determination is given in 
Section 3. A search for microlensing events is depicted in Section 4. Finally, a discussion  
appears in 
Section 5.

\section{Observations and data reduction}

A lens monitoring was performed during twelve years, 1999 to 2010, using 
the 82cm IAC80 telescope at the Instituto de Astrof\'\i sica de Canarias' Teide Observatory 
(Tenerife, Canary Islands, Spain). Two different CCD were used.
From 1999 to 2005 a Thomson 1024 x 1024 chip was employed, giving a field of view
of about 7$\farcm$5, with a pixel size of 0$\farcs$43. Since 2005 a new CCD,
CAMELOT, was installed. CAMELOT hosts a E2V 2048 x 2048 chip with 0$\farcs$304 pixels, 
corresponding to a 10.4 x 10.4 arcmin$^2$ FOV. A standard {\it R} broadband filter was 
always used for the observations, fairly closely to the Landolt system \citep{lan92}.
The final data set comprises 105 points, each of them corresponding to several 900s
exposures. 

A remarkable characteristic of the photometric data presented here is their high degree 
of homogeneity; they were obtained using the same telescope and filter over the entire 
monitoring campaign. Therefore, the reduction process can be the same for all the frames. 
In a first step, the data were reduced using standard procedures included in 
IRAF\footnote{IRAF is distributed by the National Optical Astronomy 
Observatories, which are operated by the Association of Universities for Research in 
Astronomy, Inc., under cooperative agreement with the National Science Foundation.} 
(Image Reduction and Analysis Facility, see http://www.noao.edu for more information) {\it 
ccdred} package.

We consider two main sources of error in our observations:
\begin{enumerate}
\item{{\bf On the one hand we take into account Extinction Errors:}
The main part of the variability of the observed target magnitude is explained in terms 
of atmospheric extinction and air-mass variability. Extinction errors are complicated 
by color terms when broad multi-band photometry is dealt with.} 
\item{{\bf On the other hand we consider Aperture Photometry errors:}
Due to the configuration of Q0142-100 system, there are some specific aperture photometry 
errors to take into account. As demonstrated in \citet{ser99}, these errors are driven 
by seeing variations, and can be separated in two parts as follows.
i) Influence of the lens galaxy: Since the core of the lens galaxy is very close to 
the B image, most of the galaxy's light lies inside the image B aperture, but outside 
the image A aperture. 
ii) Overlapping of images. The separation between the two images is $2\farcs2$ and 
hence, when poor seeing conditions prevail, there is an important effect of 
cross-contamination of light between the two quasar images.} 
\end{enumerate}

For extinction errors, the best and traditional method to work with is to measure 
differential photometry with several field stars close to the lens components \citep{kje92}.
Five stars in the vicinity of Q0142-100 -defined as 1,2,3,4 and 5 in \citet{nak03}- are 
used. These stars  were examined differentially in sets of 4 versus one star. This allowed 
us to establish the stability of each comparison star. After careful analysis, only two 
stars -1 and 3 for 1999-2005, 1 and 2 for 2007-2010- were selected as comparison stars for 
differential photometry. Photometric errors were calculated using 
the statistical error analysis developed by \citet{how88}, which uses the rms of the 
differential photometry of comparison stars (1-3 or 1-2 in our case) to deduce the 
photometric errors of lens components A and B. Figure 2 shows the differential R 
magnitude for comparison stars. The final results are in good concordance with the values obtained by 
\citet{nak03}, who gave $R3-R1 = 1.20 \pm 0.02$ and $R2-R1 = 0.73 \pm 0.02$.

The solution for aperture photometry errors presents a higher level of difficulty. Although 
the two quasar images appear separated in the individual frames, the lensing galaxy is too 
faint to be detected ($R$ =19.35), and therefore its light it is not considered in the final 
model fitting. Accurate photometry was obtained by simultaneously fitting a stellar 
two-dimensional profile on each lens component by means of PSF profiles derived from bright 
field stars (1,3,4,5 stars in the 1999-2005 data, and 1,2,3,4 stars in 2007-2010 data). The 
automated PHO2COM IRAF task, described in \citet{ser99}, was used. In order to check the 
feasibility of the PHO2COM method R magnitude and errors for both components are plotted 
versus FWHM in Figures 4,5,6 and 7. As demonstrated in \citet{ser99}, aperture photometry
errors (mainly due to the influence of the lens galaxy and an overlapping of the lens 
component images) are driven by seeing variations. No seeing dependences are detected in 
the photometric error (or magnitudes), as can be seen in Figures 6 and 7 (and 4 \& 5), which 
means that aperture errors are minimized. On the other hand it is also useful to show the 
final field obtained as a combination of all the individual subtracted images for 2007-2010 
data. As shown in Figure 3, whereas the lens components are eliminated (the residuals of
component A are of around 0.01\% of the original flux) some galaxy light 
remains, an additional proof that the PHO2COM method works well. 

Finally, the apparent magnitudes of the A and B components were derived by comparing the 
instrumental fluxes with the star labelled as 1 (see \citet{nak03}) in both datasets. The 
results are shown in Figures 4 and 5. The average magnitudes of the A and B components are 
16.42 and 18.42, with a standard deviation of 0.19 and 0.23, respectively, for the 1999-2005 
data, whereas for 2007-2010 the results are 16.51 and 18.50 for A and B magnitudes, with 
errors of 0.06 and 0.07, respectively.

The mean difference in magnitude among both components, $\Delta M = M_B - M_A$, is 2.00 for 
the two datasets. For the 1999-2005 data, the photometric errors (derived using the 1 and 3 
field stars) of the A component data are of the order of 0.02 mag, while those of the B 
component are around 0.2 mag (see Figure 6). On the other hand, for the 2007-2010 data the 
use of the new CCD, with a different pixel size and closer comparison stars (1 and 2), 
allows to improve the error bars of both components, 0.05 and 0.007 mag for B and A, 
respectively (see Figure 7).  

The result of the monitoring program is shown in Figure 8, where the light curves
for component A (black) and B (red) in the $R$ band are presented.

\section{Time delay}

To extract useful information from the light curves of the components of a gravitational 
lens system is required a high degree of photometric accuracy. However, Q0142-100 is a 
quite difficult system to analyse, and not only due to its very complicated configuration, 
with the underlying lensing galaxy close to the faintest component. 
Additional general drawbacks are the small variation in magnitude of the light curves 
during the whole period and, in our particular case, the large errors in the magnitude of 
component B when the IAC80's old CCD was used, the two large gaps in the data and the 
relatively small amount of data points obtained in some of the observational seasons. On 
the other hand, the system can be observed during 210-220 days each year from the Teide 
Observatory, much longer than from other observatories, which seems enough for time delay
calculations given the value obtained by \citet{kop12}

Each of our observational seasons covered several months of data, from the shortest 
last campaign of 53 days to a maximum of 201 days in the second one, with five seasons 
covering more than 125 days. On the other hand, the inter-seasons gaps -besides of the two 
large ones- are in the 150-240 days interval. These two facts mean that our database allow 
the detection of a large range of possible time delays, between 10 and 250 days.

The dataset presents two main gaps (see Figure 8): TJD-2935 to TJD-3985 and TJD-4677 to 
TJD-5305, so three different natural subsets can be selected, in which the average mags are: 
$M_A(1) = 16.54$ $(\sigma = 0.12)$, $M_B(1) = 18.55$ $(\sigma = 0.16)$, $M_A(2) = 16.33$ 
$(\sigma = 0.13)$, $M_B(2) = 18.34$ $(\sigma = 0.13)$, $M_A(3) = 16.51$ $(\sigma = 0.06)$, 
$M_B(3) = 18.50$ $(\sigma = 0.07)$. The average mags for the whole dataset are:
$M_A = 16.47$ $(\sigma = 0.14)$, $M_B = 18.47$ $(\sigma = 0.15)$. 

Prior to calculate the time delay, data must be checked to remove possible strong and 
simultaneous (not time-shifted) variations of data points in both components. These 
points probably originated from failures in the CCD or bad weather conditions, 
and their inclusion leads to artificial features in the light curves and so to wrong 
time delay determinations. To avoid this we have eliminated the points with a 
simultaneous difference in magnitude in both components larger than 2.5 times their error 
bar as compared with the previous and following records. This has been applied
to those points with a difference in their observation dates of less than 10 days. 
Only 3 of the 105 initial points had to be removed.

\subsection{The $\delta^2$ method}

There are several "classical ways" of obtaining the time delay between the components 
of a variable quasar from discrete, unevenly sampled temporary series: discrete 
correlation function \citep{ede88}, dispersion spectra \citep{pel96},
linear interpolation \citep{kun97}, z-transformed discrete correlation function
\citep{ale97}, etc. (see \citet{osc01} and references therein for a brief
depiction of all these methods). In this paper we will use the $\delta^2$ method 
\citep{ser99} to calculate the time delay of Q0142-100, as it offers very good results 
even with large gaps \citep{osc01} and small variations in the flux of the components. 

This method makes use of the similarity between the discrete
autocorrelation function (DAC) of the light curve of one of the components (A, for
example) and the A-B discrete cross-correlation function (DCF). This second order 
technique helps to improve the estimation of the time delay, as was stated in
\citet{osc01}, where a comparison of the results obtained with several statistical 
methods is made. \citet{osc01} demonstrate, with simulated and real data coming 
from different telescopes, that the $\delta^2$ method offers the best results 
without interpolating data even when large gaps are present. 

The light curves of both components of a gravitationally lensed quasar have the 
same origin, and so the same shape, which guarantees the fulfilment of the 
relationship DCF($\tau$) $\simeq$ DAC($\tau - \Delta\tau_{\rm BA}$), of course
when no strong microlensing masks the QSO's intrinsic variability. From the DAC 
and DCF functions one can define a function,
\begin{equation}
\delta^2 (\theta) = \left( {1 \over N} \right) \sum_{i = 1}^N 
S_i \left[ {\rm DCF}(\tau_i) - {\rm DAC}(\tau_i - \theta)\right]^2 \, ,
\end{equation}
for every fixed value $\theta$ (days), with $S_i = 1$ when both the DCF and DAC
are defined at $\tau_i$ and $\tau_i - \theta$, respectively, and 0 otherwise. The
most probable value of the time delay will correspond to the minimum of this
function. 

To calculate the DAC and the DCF functions the procedure described in \citet{ede88}
(see also \citet{osc97}) was used. For two discrete data sets, $a_i$ and $b_j$, the
DCF is defined as:
\begin{equation}
DCF(\tau) = {1 \over M} \, {\left( a_i - \bar a \right) \, \left( 
b_j - \bar b \right) \over {\sqrt{ \left( \sigma_a^2 - e_a^2 
\right) \, \left( \sigma_b^2 - e_b^2 \right)}}} \, ,
\end{equation}
averaging over the $M$ pairs for which $\tau - \delta \leq \Delta t_{ij} < \tau + 
\delta$, $\delta$ and $e_k$ being the bin semi-size and the measurement error 
associated with the dataset $k$, respectively, while $\sigma_k$ is its standard 
deviation. This equation straightforwardly leads to the expression for the DAC. 

\subsection{Results}

Our first test to calculate the time delay from our data consisted in checking that in 
fact it is the B component light curve which follows component A. For this, we applied 
the $\delta^2$ method to the whole dataset, with $\delta = 5$ days, taking into account 
both possibilities: A follows B and B follows A. In both cases a time delay between 10 
and 250 days was considered and the A component data were selected for the DAC 
calculations due to the lower error bars in the photometry. No clear pattern for the 
time delay was found when the component A delayed option was chosen, and the results of 
the $\delta^2$-test when component B follows component A were 30 times better. Then, 
under the assumption that component B follows component A, the   
application of the $\delta^2$-test to the DAC and DCC curves appears in Figure 9 
(normalized to its minimum value), where the minimum of the curve appears at 72 days,
corresponding to the best delay. 

To calculate the uncertainty in our estimate of the time delay we used a Monte 
Carlo algorithm. A random number generator added a variable to each point of the
dataset giving a modified value inside the range of its observational errors (see 
\citet{efr86}), thereby obtaining standard bootstrap samples. The $\delta^2$-test was then 
applied to the bootstrap samples to get the time delay. This process was repeated
10000 times, a number large enough for the results to be treated statistically. The use 
of this Monte Carlo algorithm, again for delays between 10 and 250 days, gives a 
complicated distribution of values, as it is multimodal. The results of the Monte Carlo 
procedure are shown in Figure 10, in which the number of times that each time delay is 
obtained for the whole dataset is plotted. A sharp peak appears, corresponding to a time 
delay of 72 days, with a 68\% ($1\sigma$) of the iterations giving a time delay in the 
interval 50-94 days. Then, we will consider a value of 72$\pm$22 days for the most 
probable time delay.

As was explained in previous sections, our database allows the study of time delays
between 10 and 250 days. None of our results clearly favours the 150 days 
obtained by \citet{kop10}. However, our time delay estimate, $72\pm22$ days, does
agree within uncertainties with the time delay derived by \citet{kop12}, 
$95_{-16 -29}^{+5 +14}$ days, and even more with their most probable value of $89\pm11$ 
days. 
  
 \section{Microlensing}

The light curves of the A (black) and B (red) images are represented again in Figure 11, 
this time delaying the B data in 72 days and shifting them by $-$2.00 magnitudes, the 
result of $<M_A>-<M_B>$. As can be seen, both curves follow a similar trend, mainly 
for the last set of data where the error bars of component B are lower. However, when 
these curves are inspected in more detail it seems that, between roughly half of the 
second observational season, TJD-2800, and the end of the third season, TJD-4039, 
component B delayed and shifted data are systematically fainter than component A data.

This effect is better observed when the data corresponding to each season are grouped
and averaged. The second season, the one with more observations and when this difference 
seems to start, is divided into two. The results appear in Table 1. While the values for
most of the seasons (1, 2-first half and 4-9) are quite homogeneous, the second half of
season 2 and season 3 are significantly different. In fact, $<<M_A>-<M_B>> = -2.10$ for 
these one and half seasons and $-$1.99 for the rest of the data set. This is represented
in Figure 12. In the upper panel the seasonal averaged data are displayed, black points
for the A component and blue and red points for the B component and $-1.99$-mag shifted
B component, respectively. Notice the coincidence of the black and red points
except for these one and half season. The average data are shown in the lower panel,
with the dashed line representing $-$1.99.
 
The variation in the difference of the average magnitude of both components during 
$\sim$1200 days suggests the presence of a possible microlensing event, with our
observations corresponding to the entrance and exit of the event. Although the season 3 
points for component A do not overlap in time with the shifted season 3 points for 
component B and the discrepancy could also be caused by a decrease in the intrinsic flux 
of the quasar, it seems that microlensing is indeed the most possible explanation for 
this behaviour. Unfortunately, 
we lack data for most of this period and hence further evidence of this event can not be 
obtained. 

\section{Discussion} 

The result of a photometric follow-up of the gravitational lens system Q0142-100 in the 
$R$ band is presented in this paper. The observations, taken with the 82 cm IAC-80 
telescope, at Teide Observatory, Spain, were made from 1999 to 2010, with 105 points
obtained, as part of an on-going lens monitoring program. A calculation of the time 
delay between both components by using the $\delta^2$-test has been performed. The 
resulting delay is of 72$\pm$22 days, very different from that obtained by \citet{kop10}
but within the errorbar of a later result given by \citet{kop12}, $95_{-16 -29}^{+5 +14}$ 
days (68 and 95\% confidence intervals) and $89\pm11$ days as their most 
probable result.

\citet{leh00} select four models to fit several gravitational lens
systems, among which is Q0142-100: i) a dark matter model (SIE), ii) a model based on
photometric fits (constant $M/L$), iii) dark matter model in an external shear field
(SIE + $\gamma$), iv) photometric model in an external shear field ($M/L + \gamma$).
However, neither an $M/L$ nor a SIE model are good solutions for Q0142-100.
In the first case, a poor fit to the image positions and magnifications is obtained, whilst
a great degree of misalignment relative to the luminosity is required to obtain a good 
fit for the SIE model. Both models improve their results when an external shear of 
$\gamma \sim 0.07$ is added, but this shear does not correspond to any shear estimate 
for the nearby galaxies.

The time delays predicted by \citet{leh00} for Q0142-100 for the four different models
are, given as $h\Delta t$: SIE = 80.1$\pm$0.3, SIE + $\gamma$ = 84-87, M/L = 121.3, 
and M/L + $\gamma$ = 115$\pm$3. All of them are larger than the time delay we have
derived and even larger than the results given by \citet{kop12}. 

We have attempted to use the time delay to fit the lens models but unfortunately, 
given the large error bar, our current time delay estimation cannot be yet used as 
an extra constraint to clarify the properties of the lens mass model. 

Our estimate of the time delay and the one derived by \citet{kop12} are 
below the time delay predicted by the theoretical models, even taking into account the large
uncertainties. This could be explained if some nearby, not detected yet (maybe the 
bright galaxy just to the North of the lens system, see figures 1 and 3), components of the 
system are missed and not included into the models. These missing components can be located 
either around the lens galaxy
or on the line of sight to the quasar. Although a more accurate value of the time delay
could help to reduce the uncertainties in the model, it seems clear that finding more
details on the system environment will help even more to its understanding and reconcile
the values of the time delay with those of the Hubble constant derived by other methods.

\acknowledgments

We are especially grateful to the Instituto de Astrof\'\i sica de Canarias' support 
astronomer and night assistant teams for their help in the observations of most of 
the data appearing in this paper during the routine and service time.

The 0.82m IAC80 Telescope is operated on the island of Tenerife by the Instituto de 
Astrof\'\i sica de Canarias in the Spanish Observatorio del Teide

{\it Facilities:} \facility{IAC80 (CAMELOT)}.

\clearpage

\begin{figure}
\includegraphics[angle=0,scale=.85]{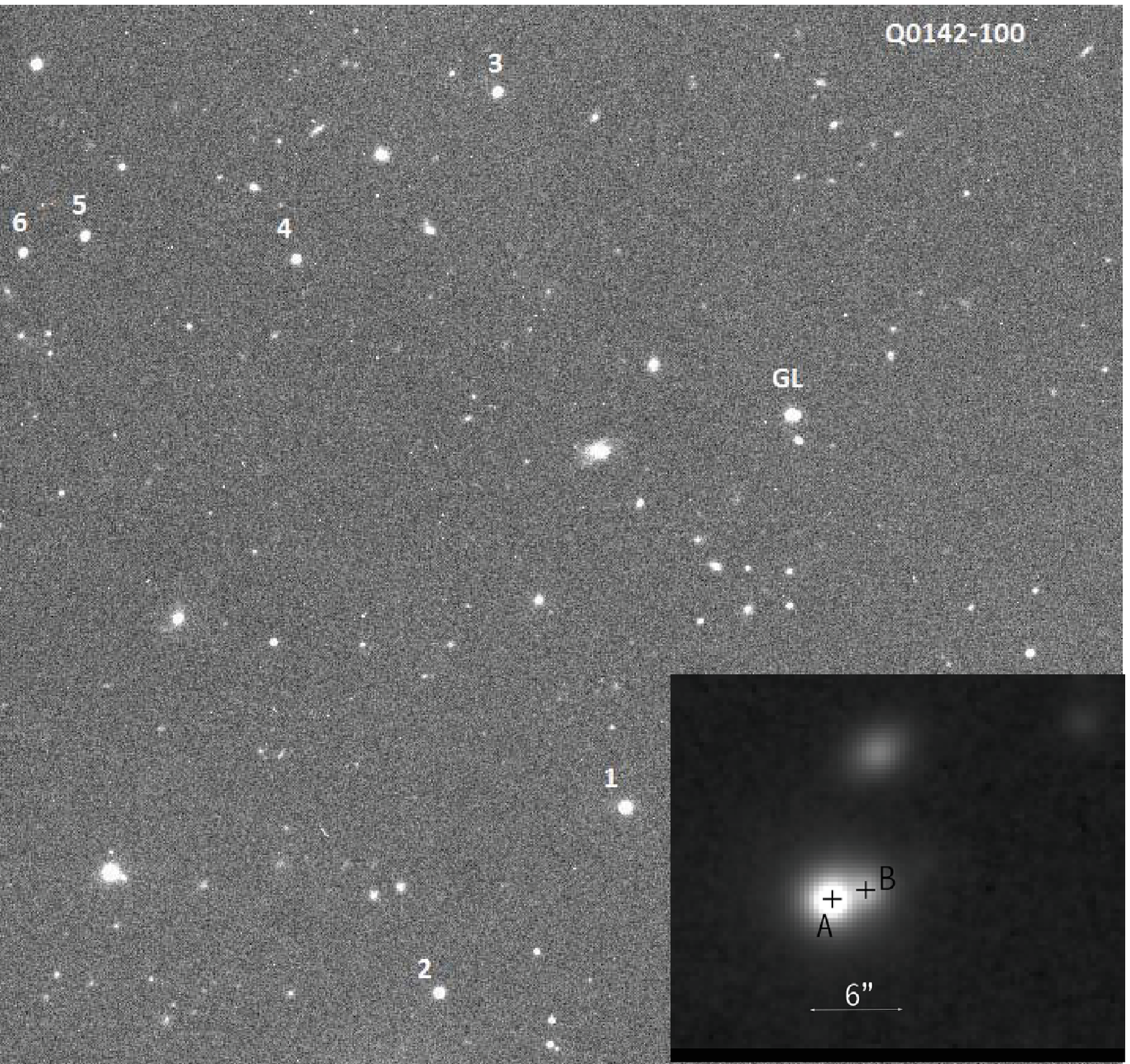}
\caption{{\it R}-band image taken with CAMELOT@IAC80 (see text) of the field surrounding 
Q0142-100 (labelled as GL) where the main reference stars are marked as 1-6 following the 
notation by \citet{nak03}. North is up and East is to the right, with a field of view of 
10.4x10.4 arcsec. A zoomed view of the lens system has been included in the lower right 
corner of the image.
\label{fig1}}
\end{figure}

\clearpage

\begin{figure}
\includegraphics[angle=270,scale=.50]{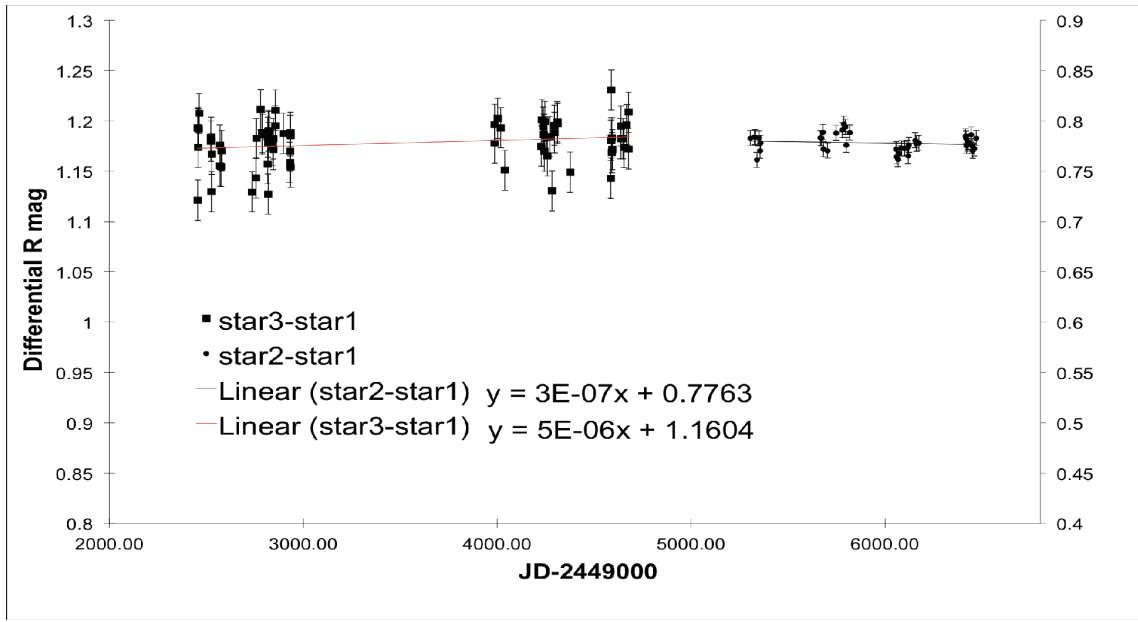}
\caption{{\it R}-band differential light curves for comparison stars. As explained in the 
text photometric errors of lens components were calculated using the rms of the differential 
photometry of comparison stars (1-3 or 1-2 in our case). The differential R magnitude for 
reference stars is 
in good concordance with the values ($R3-R1 = 1.20 \pm 0.02$ and $R2-R1 = 0.73 \pm 0.02$) 
given by \citet{nak03}. 
\label{fig2}}
\end{figure}

\clearpage

\begin{figure}
\epsscale{.80}
\plotone{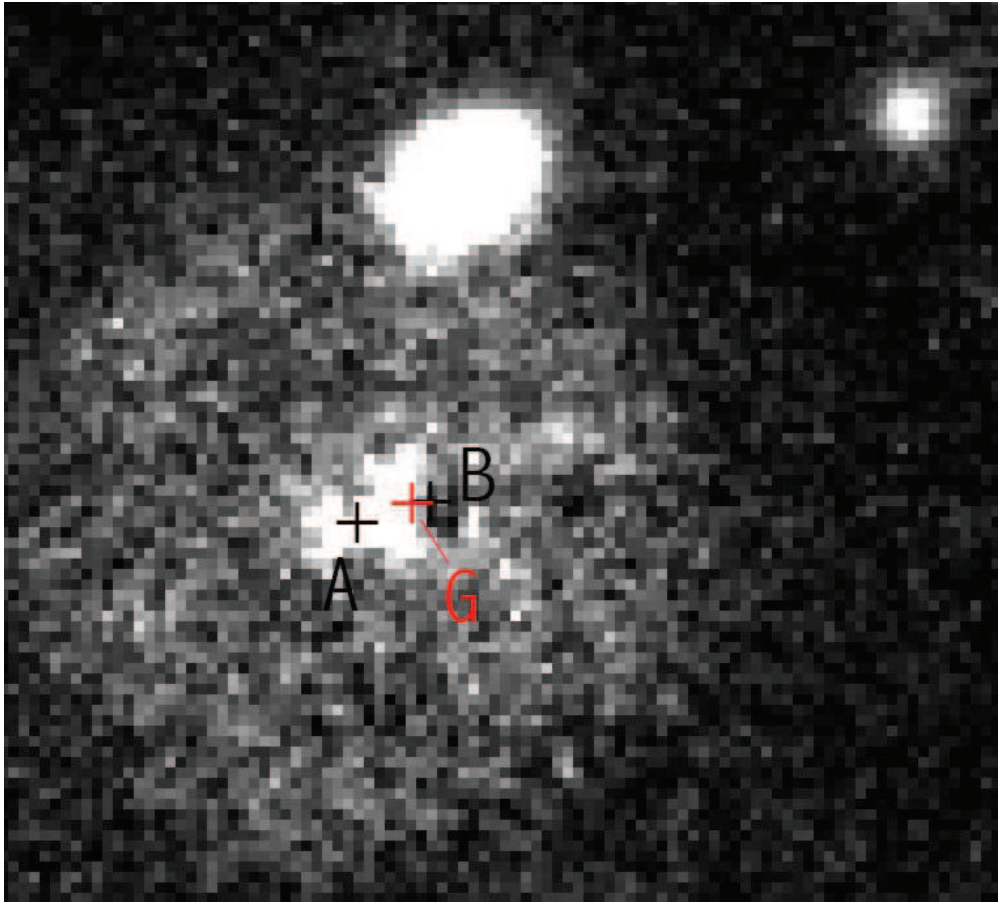}
\caption{Q0142-100 final field obtained as a combination of all the individual
subtracted images for 2007-2010 data. Lens component (A,B) and lensing galaxy (G) position 
are marked (\citet{leh00}). Whereas lens components are eliminated (the residuals of
component A are of around 0.01\% of the original flux)
some galaxy light remains, which is an additional proof that PHO2COM method is working well 
(see text for details). North is down and East is to the right, with a field of view of 
around 30 arcsec.
\label{fig3}}
\end{figure}

\clearpage

\begin{figure}
\includegraphics[angle=270,scale=.60]{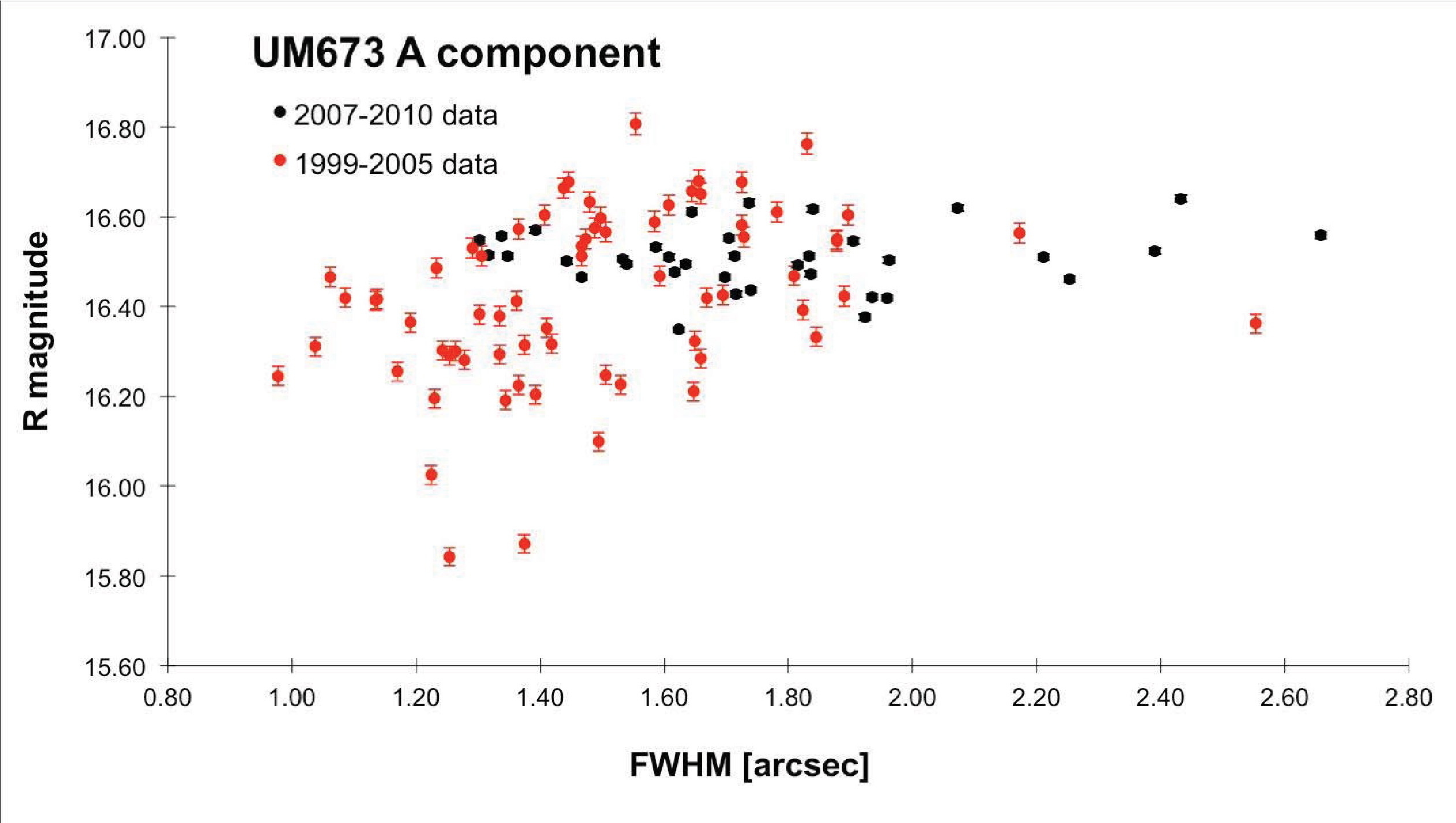}
\caption{Apparent magnitudes of the A component for the Old (red points) and New (black 
points) CCD once compared the instrumental fluxes with those of the reference star 1.
\label{fig4}}
\end{figure}

\clearpage

\begin{figure}
\includegraphics[angle=270,scale=.60]{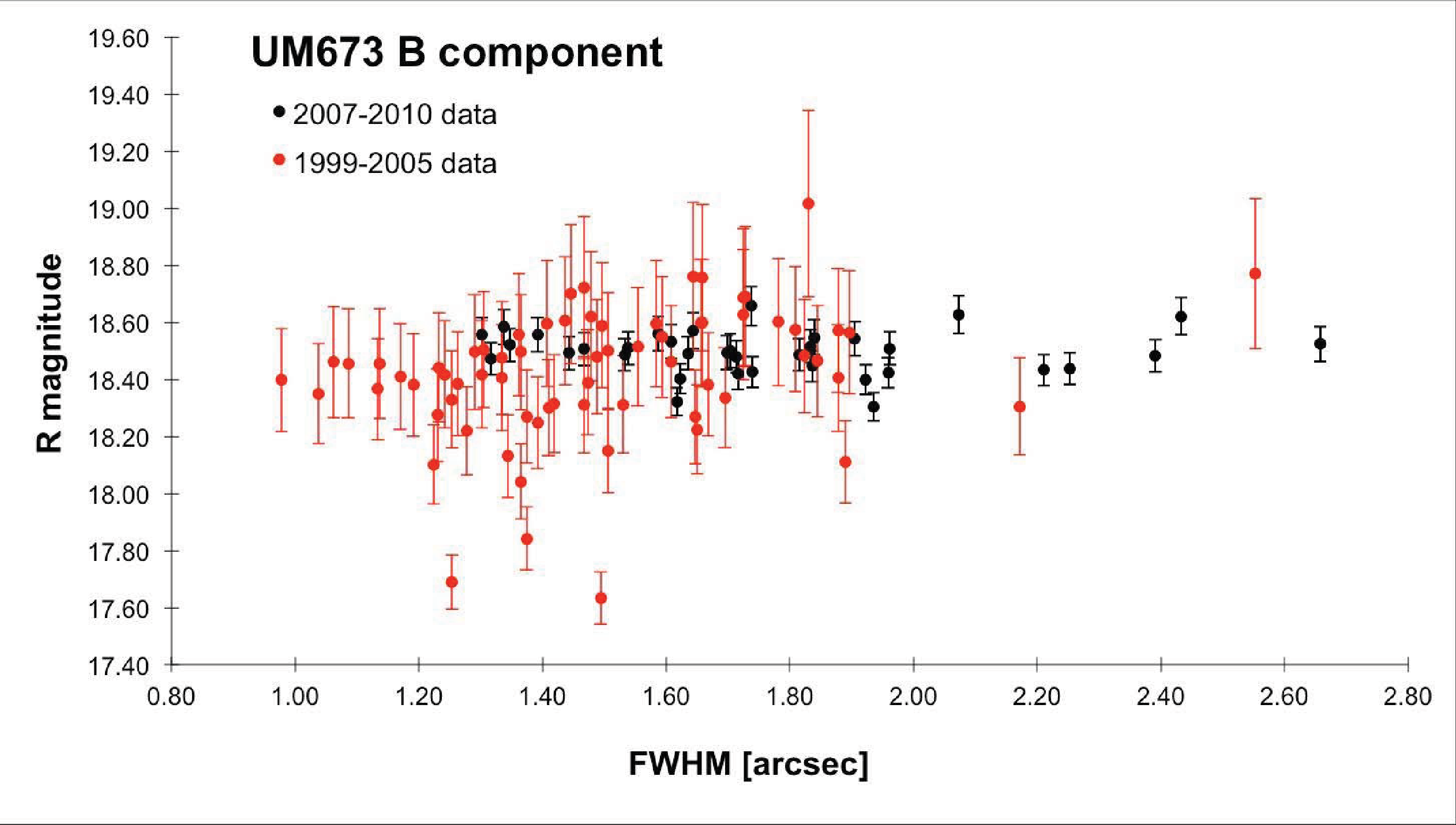}
\caption{Same as Figure 4 for the B component of Q0142-100.
\label{fig5}}
\end{figure}

\clearpage

\begin{figure}
\includegraphics[angle=270,scale=.60]{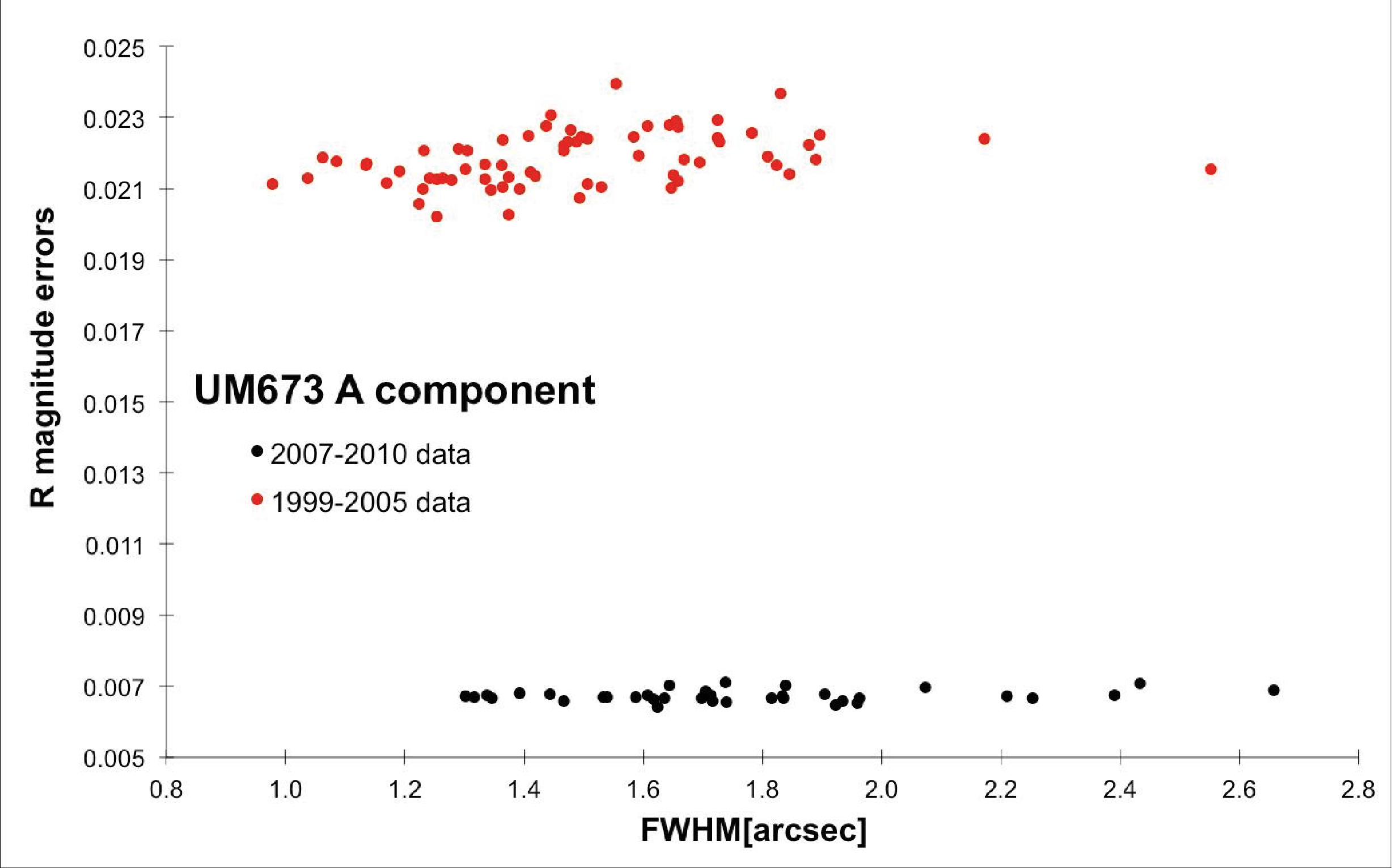}
\caption{Photometric errors of the A component of Q0142-100 for the Old (red points), derived using the 1 and 3 field stars, and New (black points) CCD, using the 1 and 2 field stars. Notice the large improvement in
the error bars when the new CCD is used. No seeing dependence is detected.
\label{fig6}}
\end{figure}

\clearpage

\begin{figure}
\includegraphics[angle=270,scale=.60]{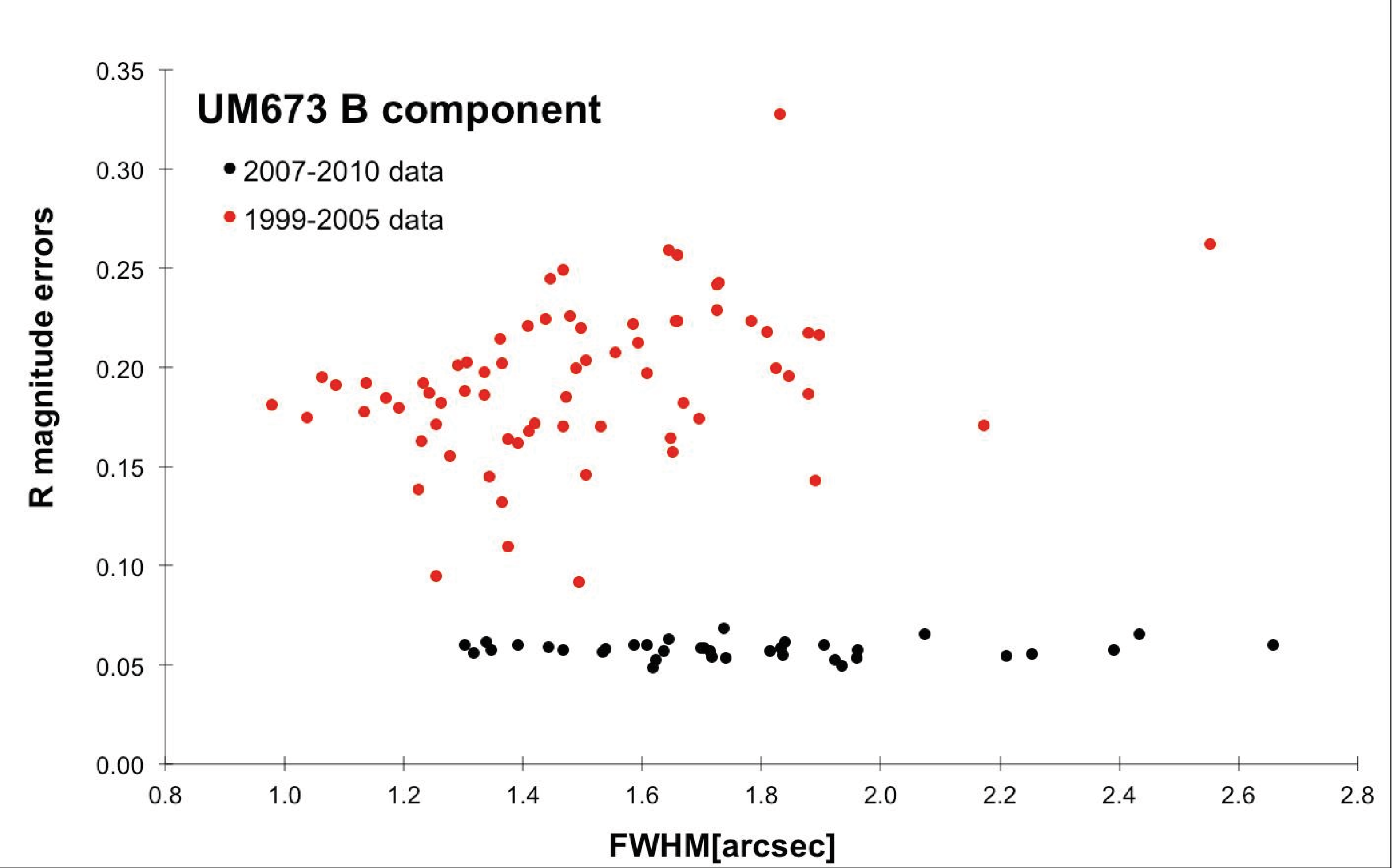}
\caption{Same as Figure 6 for the B component.
\label{fig7}}
\end{figure}

\clearpage

\begin{figure}
\plotone{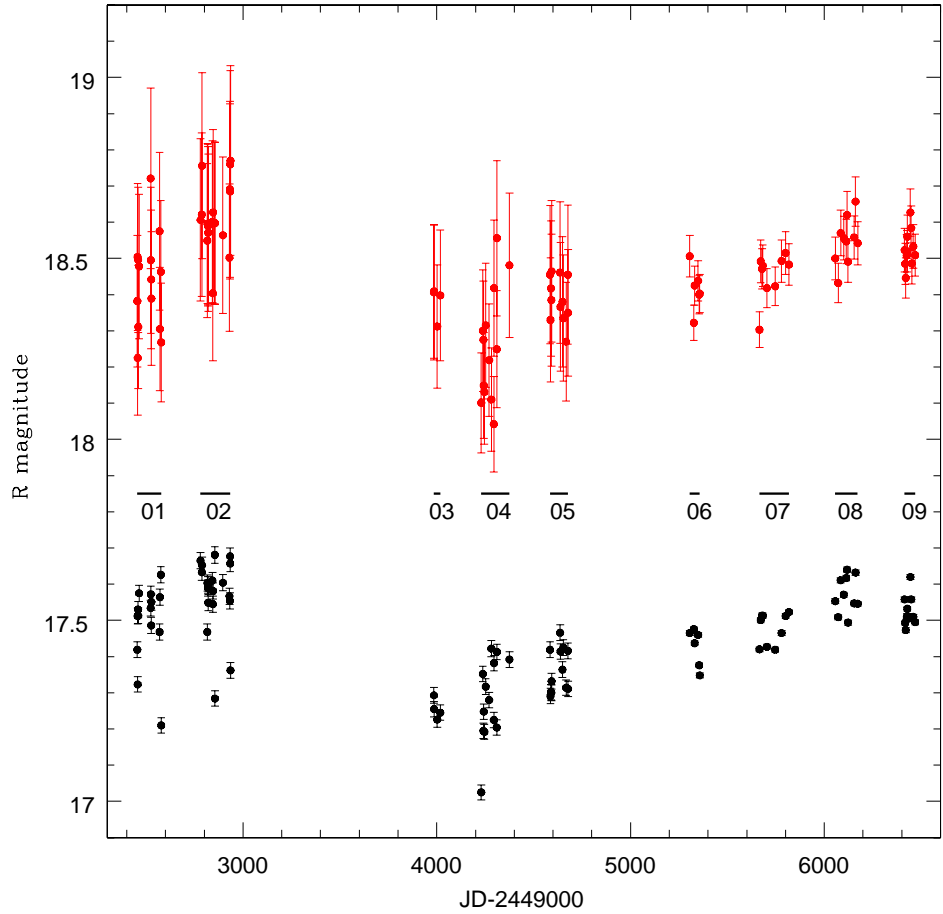}
\caption{Light curves of component A (black) and B (red) from the observations
made between October 1999 and September 2010 at the IAC80 telescope. The A component data 
has been shifted by +1 mag. The horizontal lines between both components represent the 
length of each observational season.
\label{fig8}}
\end{figure}

\clearpage

\begin{figure}
\plotone{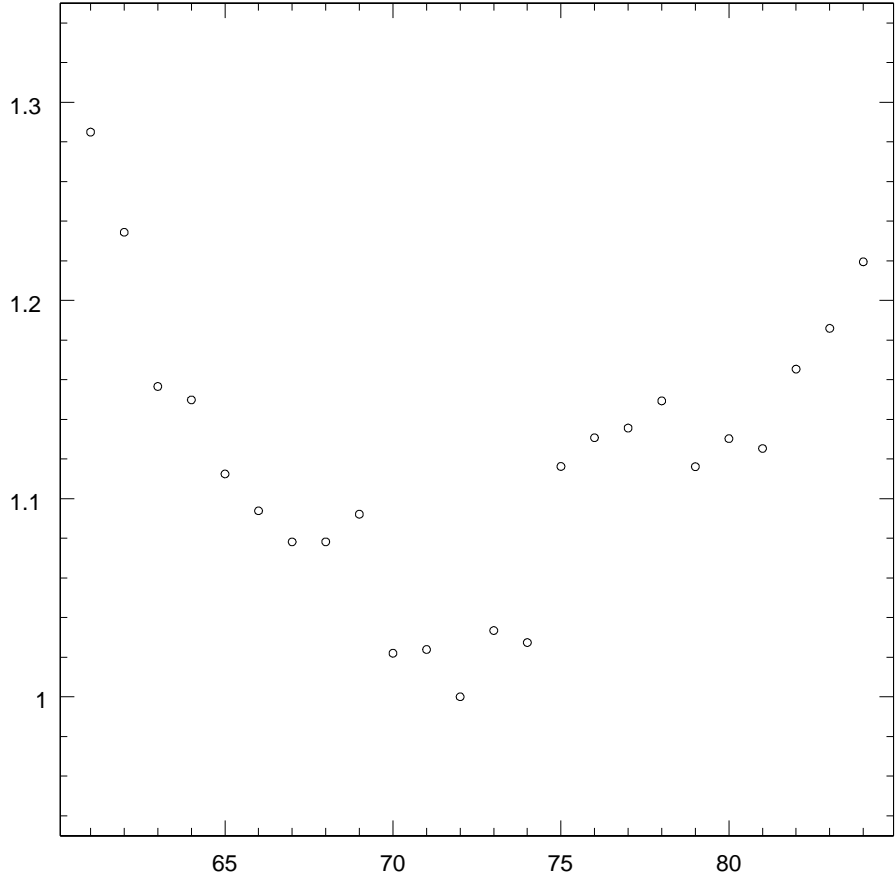}
\caption{Application of the $\delta^{2}$ test to observational data, normalized by its 
minimum value. The best time delay obtained is 72 days.
\label{fig9}}
\end{figure}

\clearpage
\begin{figure}
\plotone{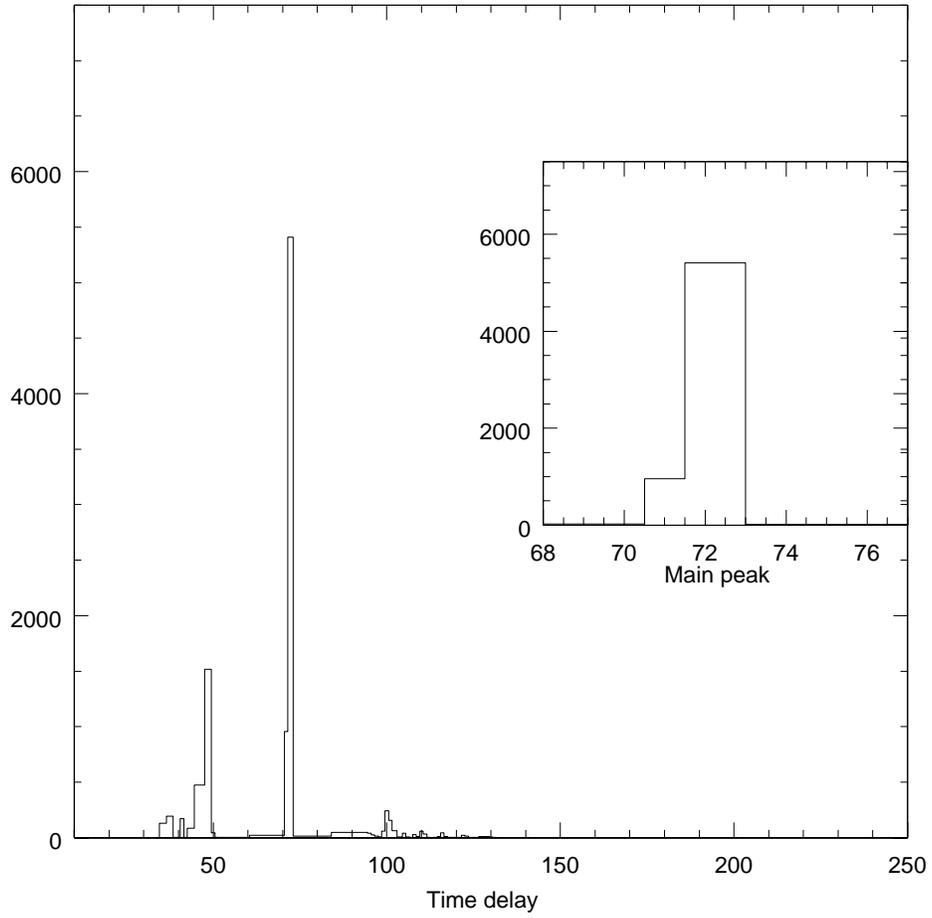}
\caption{Number of times that each time delay appears when the Monte Carlo algorithm
is applied to the whole dataset.
\label{fig10}}
\end{figure}

\clearpage

\begin{figure}
\plotone{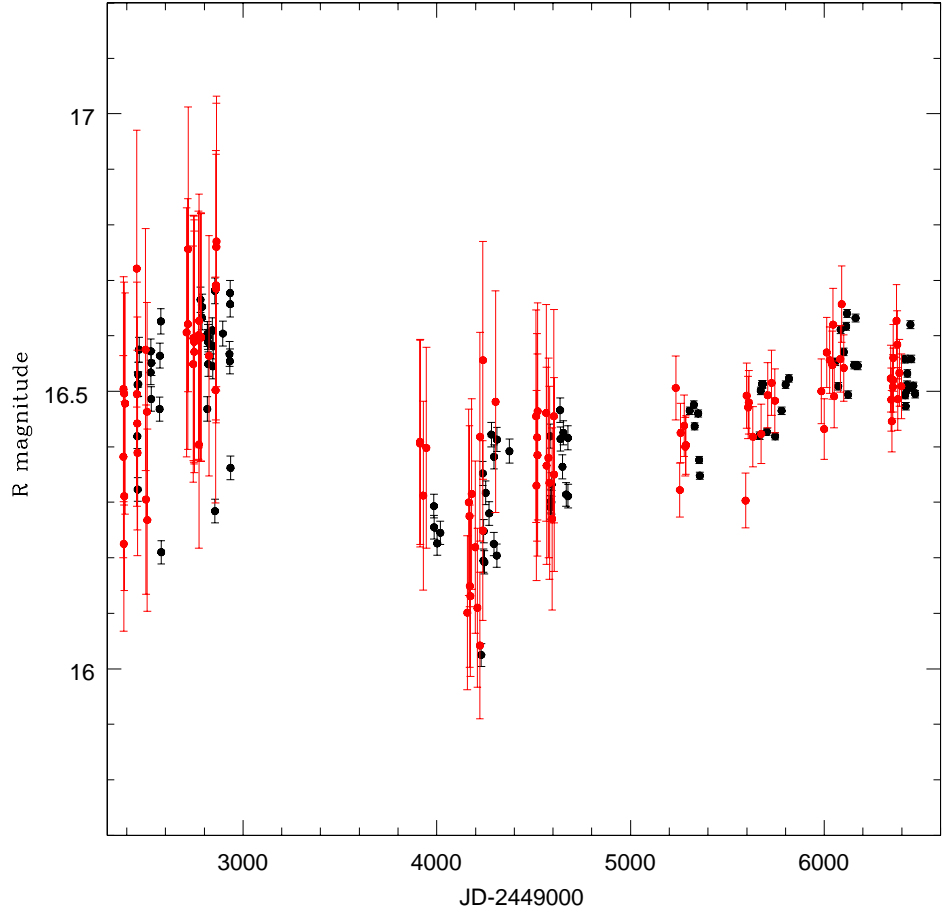}
\caption{Light curves of components A (black) and B (red) of Q0142-100. The B
component data have been shifted by 72 days, the time delay derived (see text), and
 $-$2.00 magnitudes. 
\label{fig11}}
\end{figure}

\clearpage

\begin{figure}
\plotone{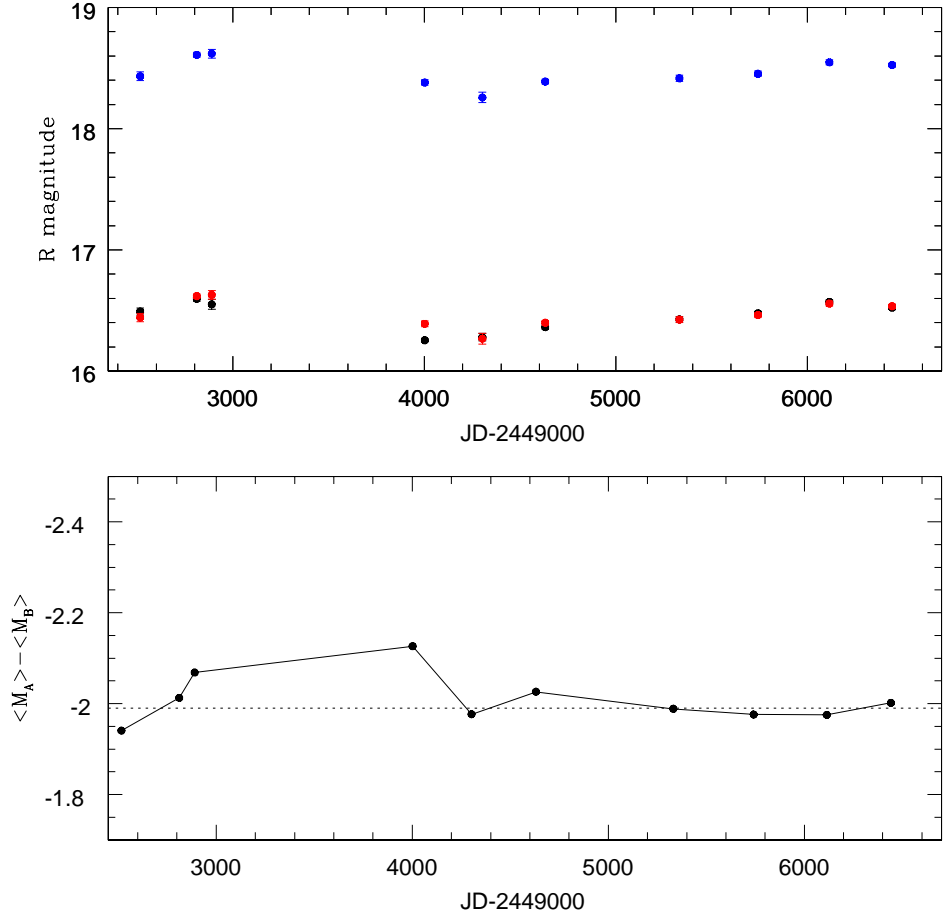}
\caption{Upper panel: seasonal averaged data, taken from Table 1, for the A (black) and B 
(blue) component. Red points correspond to the B component data shifted by $-1.99$ mags. 
In both cases the B component data have been shifted by 72 days. The errorbars are also 
included.
Lower panel: $<M_A>-<M_B>$ for each observational season. The dashed line represents the 
average value of most of the seasons, 1.99. \label{fig12}}
\end{figure}

\clearpage

\begin{deluxetable}{cccccccccc}
\tablecaption{Average values for each observational season. SD means Starting Date of the 
season, ED is the End Date, NE the Number of Epochs included, 
$E_A = \sqrt{\frac{\Sigma(x_A-<x_A>)^{2}}{n (n-1))}}$, $E_B = \sqrt{\frac{\Sigma(x_B-<x_B>)^{2}}{n (n-1))}}$, 
1H is the first half of season 2 and 2H is the second half.}
\tablewidth{0pt}
\tablehead{\colhead{Season}&\colhead{$<TJD>$}&\colhead{SD}&\colhead{ED}
&\colhead{NE}&\colhead{$<M_A>$}&\colhead{$<M_B>$}&\colhead{$E_A$}&\colhead{$E_B$}
&\colhead{$<M_A>$-$<M_B>$}}
\startdata
1&2516.50&2454.64&2578.36&14&16.49&18.43& 0.03 & 0.03 & $-$1.94 \\
2 (1H)&2811.62&2780.71&2842.53&9&16.60&18.61& 0.02 & 0.02 & $-$2.01 \\
2 (2H)&2889.98&2844.60&2935.37&10&16.55&18.62& 0.04 & 0.04 & $-$2.07 \\
3&4001.92&3985.45&4018.38&4&16.25&18.38& 0.01 & 0.02 & $-$2.13 \\
4&4302.53&4229.71&4375.34&13&16.28&18.26& 0.03 & 0.04 & $-$1.98 \\
5&4631.05&4584.71&4677.39&12&16.36&18.39& 0.02 & 0.02 & $-$2.03 \\
6&5331.62&5305.69&5357.54&6&16.43&18.42& 0.02 & 0.02 & $-$1.99 \\
7&5741.58&5665.72&5817.44&9&16.48&18.45& 0.01 & 0.02 & $-$1.98 \\
8&6114.56&6056.62&6172.5&10&16.57&18.55& 0.02 & 0.02 & $-$1.98 \\
9&6442.00&6414.50&6469.5&11&16.52&18.53& 0.01 & 0.02 & $-$2.00 \\
\enddata
\end{deluxetable}

\end{document}